\documentclass[aps,prl,showpacs,amsmath,amssymb,amsfonts,twocolumn,
superscriptaddress,floatfix,footinbib]{revtex4}

\usepackage{subfigure}
\usepackage{graphicx}
\usepackage[flushmargin]{footmisc}
\usepackage{color}
\usepackage{soul}

\newcommand{\kk}{\mathbf k}
\newcommand{\kp}{\mathbf k'}

\newcommand{\Ham}{\hat{\mathbf H}_{ij}}

\begin{document}

\title{Anisotropic Optical Spin Hall Effect in Semiconductor Microcavities}

  \author{A Amo}
    \affiliation{Laboratoire Kastler Brossel, Universit\'{e} Pierre et Marie Curie,
    Ecole Normale Sup\'{e}rieure et CNRS, UPMC Case 74, 4 place Jussieu,
    75252 Paris Cedex 05, France}

  \author{T C H Liew}
    \affiliation{Centre for Quantum Technologies, National University of Singapore, Singapore 117543}

  \author{C Adrados}
    \affiliation{Laboratoire Kastler Brossel, Universit\'{e} Pierre et Marie Curie,
    Ecole Normale Sup\'{e}rieure et CNRS, UPMC Case 74, 4 place Jussieu,
    75252 Paris Cedex 05, France}

  \author{E Giacobino}
    \affiliation{Laboratoire Kastler Brossel, Universit\'{e} Pierre et Marie Curie,
    Ecole Normale Sup\'{e}rieure et CNRS, UPMC Case 74, 4 place Jussieu,
    75252 Paris Cedex 05, France}

  \author{A V Kavokin}
    \affiliation{School of Physics and Astronomy, University of
    Southampton, Highfield, Southampton SO17 1BJ, UK}
    \affiliation{Marie-Curie Chair of Excellence ``Polariton devices", University of Rome II, 1, via della Ricerca Scientifica, Rome, 00133, Italy}

  \author{A Bramati}
    \affiliation{Laboratoire Kastler Brossel, Universit\'{e} Pierre et Marie Curie,
    Ecole Normale Sup\'{e}rieure et CNRS, UPMC Case 74, 4 place Jussieu,
    75252 Paris Cedex 05, France}


\date{\today}

\begin{abstract}

\noindent Propagating, directionally dependent, polarized
spin-currents are created in an anisotropic planar semiconductor
microcavity, via Rayleigh scattering of optically injected polaritons
in the optical spin Hall regime. The influence of anisotropy results in the suppression or enhancement of the pseudospin precession of polaritons scattered into different directions. This is exploited to create intense spin currents by excitation on top of localized defects. A theoretical model considering the influence of the total effective magnetic field on the polariton pseudospin quantitatively reproduces the experimental observations.
\end{abstract}

\maketitle

\emph{Introduction.-} The field of spintronics
has come a long way since the study of magnetic dependent electron
transport~\cite{Tedrow1971,Julliere1975} and magnetic injection of
electron spins~\cite{Johnson1985} that began in the 1970s. The spin of an ensemble of particles, such as electrons, can
encode a 
 bit of information and today this is exploited to
produce non-volatile memory elements with high read/write speeds,
high density and low power consumption~\cite{Gerrits2002}.

Semiconductors have the potential for lower operating power and
smaller size than their metallic counterparts, and semiconductor
spintronics~\cite{Awschalom2007} has been developing in parallel
with metallic spintronics, however with some difficulties for the spin injection~\cite{
Hanbicki2003
}. An alternative
method for the creation of spin currents in semiconductors is provided by
the spin Hall effect
(SHE)~\cite{Dyakonov1971,Kato2004,Wundrlich2005}, which leads to a
separation of spin-up and spin-down polarized electrons in both real
and momentum space. 
 However, this technique presents the problem of the rapid spin decay and dephasing due to the strong carrier-carrier scattering.

Recently, spin currents were generated in a different system,
a semiconductor microcavity, using the so-called optical spin Hall
effect (OSHE)~\cite{Leyder2007}. This optical analogue of the SHE
created separated spin currents of polaritons in real and momentum space that traveled a distance of the order of
$100\mu m$, evidencing the realistic potential applications of polariton spin on the implementation of integrated polarization based optical gates~\cite{Liew2008}. In this letter we show that we can use anisotropic fields to efficiently generate such spin currents in a preferred direction at engineered points of the sample.

Semiconductor microcavities are planar nanostructures
designed to strongly couple light and matter; a pair of highly reflective mirrors confines cavity photons which are resonant
with the exciton of a quantum well embedded in the cavity. The normal modes of such a
system are exciton-polaritons, neutral particles formed from the superposition of excitons and photons 
~\cite{Kavokin2007}, which do not suffer from the fast dephasing experienced by electrons.

The OSHE originates from the combined influence of the polarization splitting of
transverse electric and transverse magnetic modes (TE-TM splitting)
and Rayleigh scattering on polaritons that are excited by an optical
pump~\cite{Kavokin2005}. The TE-TM energy splitting arises from the dependence of the polariton energy on the angle between the polariton momentum $\kk$, and its dipole oscillation direction, determined by the plane of polarization of the linearly polarized excitation. The role of the Rayleigh scattering is to provide the system with polaritons with a pletora of different momenta after pumping with a given~$\kk$. In a generalized form, the effect of the TE-TM splitting on the polarization of exciton-polaritons can be conveniently described within the pseudospin formalism by the introduction of an effective magnetic field whose orientation depends on the polariton wave-vector. The pseudospin depicts the polariton polarization state, while the effective field is an optical analogue of the Rashba-field acting upon electron spins in semiconductor crystals lacking inversion symmetry~\cite{Dyakonov2007}. The effective field results in a $\kk$-dependent precession of the polariton pseudospin, which leads to the separation of spin up and spin down polaritons in the Rayleigh ring~\cite{Leyder2007}.

In realistic microcavities, the TE-TM splitting competes with the polarization splitting generated by the in-plane anisotropy of the sample, caused by interface fluctuations, photonic disorder or local strains~\cite{Klopotowski2006, Krizhanovskii2006,Kasprzak2006,Glazov2007}.
Anisotropy results in the appearance of a wave-vector independent linear polarization splitting equivalent to a supplementary uniform effective magnetic field acting upon the polariton pseudospins. In this
letter we experimentally study the polarization dynamics of exciton-polaritons in the presence of such a field and of the TE-TM splitting, and demonstrate the appearance of the anisotropic OSHE.
Furthermore, the anisotropic field enhances the spin precession of polaritons moving along directions given by the field orientation. This is used to efficiently generate spin currents at
engineered points in the sample, where a localized photonic defect, creating an isolated peak in the
polariton potential, is present. Our results are in agreement with a theoretical model based on the spinor Schr\"{o}dinger equation, which accounts for both the TE-TM splitting and anisotropy effective fields.

\begin{figure}[h]
        \centering
        \includegraphics[width=8.116cm]{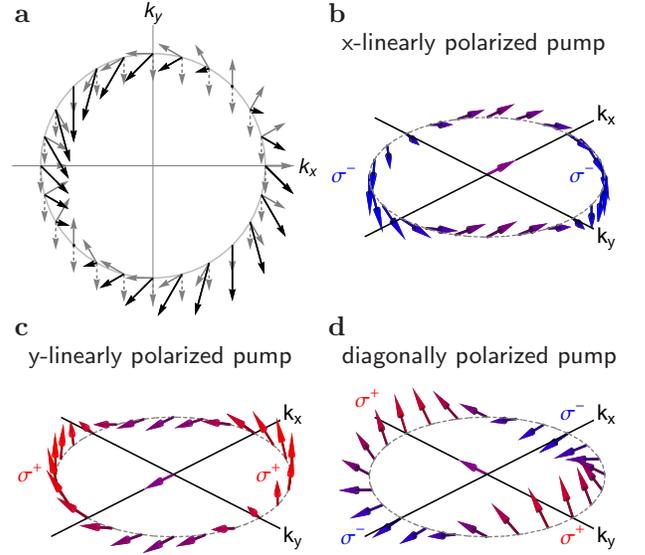}
    \caption{(color online) (a) The TE-TM splitting (grey arrows) and polarization anisotropy (grey dashed arrows) can be represented by effective magnetic fields, which act on the polariton pseudospin vector (Stokes vector). The total effective magnetic field (black arrows) causes the polariton pseudospin vectors to precess. Panels (b-d) illustrate the expected evolution of the pseudospin of polaritons that have elastically (Rayleigh) scattered from the pump state (at $\kk=0$) to a ring in reciprocal space with different planes of linear polarization of excitation. The purple arrow in the center depicts the initial pseudospin. Pseudospins pointing out of the $k_z=0$ plane give rise to circularly polarized emission.}
    \label{fig:Pseudospin}
    \end{figure}

\emph{Theory.-} Given their long decoherence time~\cite{Langbein2007},
polaritons can be described by a two-component wavefunction,
$\psi_i(\kk)$, where $i$ represents either spin-up ($+$) or
spin-down ($-$) polaritons (here we consider only lower branch
polaritons). These two spin components are directly coupled to the
circular polarizations of external light, allowing for their optical
excitation and polarization resolved detection. The two-dimensional
(spinor) wavefunction obeys a Schr\"{o}dinger equation, which can be written~\cite{Liew2009}:
\begin{align}
i \hbar \frac{\partial \psi_i(\kk)}{\partial
t}&=\Ham(\kk)\psi_j(\kk)+\int{V(\kp)\psi_i(\kk-\kp)d\kp}\notag\\
&\hspace{10mm}+f_i(\kk)-\frac{i\hbar}{2\tau(\kk)}\psi_i(\kk).
\label{eq:Shrodinger}
\end{align}
\noindent The Hamiltonian can be written:
\begin{equation}
\Ham(\kk)=\delta_{ij}T(\kk)+\frac{\hbar}{2}\left(
{\mathbf\sigma.\Omega}(\kk)\right)_{ij},
\end{equation}
\noindent where $T(\kk)$ represents the (non-parabolic dispersion)
of lower branch polaritons, ${\mathbf\sigma}$ is the Pauli matrix
vector and ${\mathbf\Omega}$ is an effective magnetic field.
${\mathbf\Omega}$ is given as the sum of a field representing the
TE-TM splitting~\cite{Kavokin2005} (this field depends on the
direction of the polariton wavevector, $\kk$) and an effective field arising from the anisotropy in the microcavity heterostructure ${\mathbf\Omega_{an}}$ (this field is wavevector independent). These fields
are illustrated in Fig.~\ref{fig:Pseudospin}a.

\begin{figure}[t]
        \includegraphics[width=8.116cm]{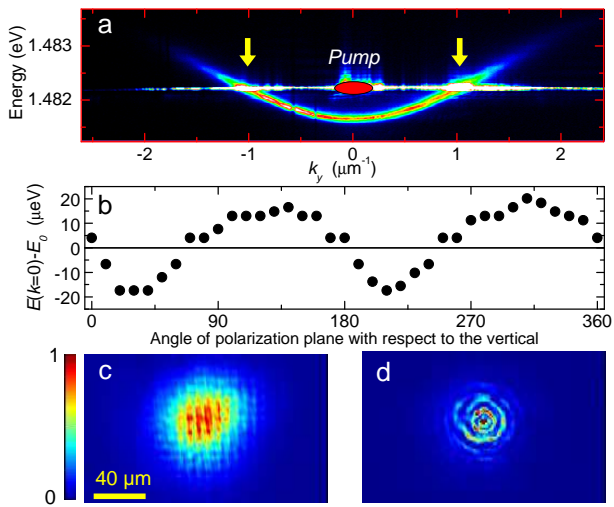}
    \caption{(color online) (a) Lower polariton branch (LPB) luminescence under out of resonance excitation at $\kk=0$ (indicated by the red spot). Resonant Rayleigh scattering takes place to LPB states marked by arrows. (b) Energy of the photoluminescence peak at $\kk=0$ under circularly polarized excitation as a function of the angle of the detected linear polarization plane ($E_0=1.487572~eV$). (c) Real space transmitted intensity in the presence of the shallow disorder usual in standard microcavities, and (d) in presence of a localized photonic defect in the center of the spot. Each panel has been normalized to show the best visibility. }
    \label{fig:dispsplit}
    \end{figure}

$V(\kk)$ represents the polariton potential. We will consider two
kinds of profile - one representing the disorder, which is generated
by a stochastic field~\cite{Savona2006} characterized by some
correlation length $l$ and a root mean squared amplitude $W_{rms}$;
and one representing an isolated peak. The continuous wave optical
pump is represented by:
\begin{equation}
f_i(\kk)=A_i e^{-\kk^2L^2/4}\frac{i\Gamma e^{-i E_p
t/\hbar}}{T(\kk)-E_p-i\Gamma},
\end{equation}
\noindent where the pump is Gaussian shaped with energy $E_p$,
linewidth $\Gamma$, spot-size $L$ and amplitude and polarization
given by $A_i$. Unlike the experiments in
Refs.~\cite{Leyder2007,Liew2009} we are working with a pump oriented
at normal incidence (which excites polaritons with zero in-plane
wavevector, $\kk=0$). Provided the pump energy is tuned above the energy
$T(0)$, Rayleigh scattering still allows the excitation of a ring in
reciprocal space (see Fig.~\ref{fig:dispsplit}a), which is important for the OSHE. Note that
Rayleigh scattering alone preserves the polariton polarization.

The decay of polaritons caused by exciton recombination and the
escape of photons through the Bragg-mirrors of the microcavity is
modeled phenomenologically with a lifetime, $\tau(\kk)$, which can
be related to the exciton and photon lifetimes, $\tau_X$ and
$\tau_C$ respectively, by
$1/\tau(\kk)=|X(\kk)|^2/\tau_X+|C(\kk)|^2/\tau_C$,
where $X(\kk)$ and $C(\kk)$ are the Hopfield
coefficients~\cite{HaugKoch,Laussy2004}.

The reason for defining the effective magnetic field
${\mathbf\Omega}$ is that it allows a qualitative understanding of
the combined effects of TE-TM splitting and anisotropy. The
polarization state of polaritons can be represented by the
pseudospin vector ${\mathbf
\rho}=(S_x/S_0,S_y/S_0,S_z/S_0)$, equivalent
to the Stokes vector for light, where
$S_x=\psi_+^*\psi_-+\psi_-^*\psi_+$,
$S_y=i(\psi_-^*\psi_+-\psi_+^*\psi_-)$, $S_z=|\psi_+|^2-|\psi_-|^2$
and $S_0=|\psi_+|^2+|\psi_-|^2$. Graphically, in the Poincar\'e sphere, ${S_x}$ and ${S_y}$ are the pseudospin coordinates laying on the equatorial plane, which represent the degree and plane of the linearly polarized component of the polarization, while ${S_z}$ describes its circularly polarized component~\cite{Leyder2007}.
 The field ${\mathbf\Omega}$ causes
the pseudospin vector to precess, giving rise to the appearance of circularly polarized emission even under linearly polarized excitation. Figures~\ref{fig:Pseudospin}b-d illustrate the pseudospin precession for different linear polarizations of the pump.

\emph{Experimental results.-} The sample used in our experiments is a $2~\lambda$, InGaAs based microcavity with a Rabi splitting of 5.1~meV. All our experiments were performed at 5~K, at zero exciton-photon detuning, using a linearly polarized single mode laser as the excitation source on a spot of 48$\mu m$ in diameter. Real and momentum space images of the emission (circularly polarized resolved) were collected in transmission geometry in two high-definition CCD cameras. The size and direction of ${\mathbf\Omega_{an}}$ has been obtained by analyzing the linear-polarization resolved photoluminescence emission of the LPB at $\kk=0$, where the TE-TM splitting vanishes. A splitting due to the anisotropy-related effective field ${\mathbf\Omega_{an}}$ of about 0.04~meV can be observed, as depicted in Fig.~\ref{fig:dispsplit}b.

    \begin{figure}[t]
        \centering
        \includegraphics[width=8.116cm]{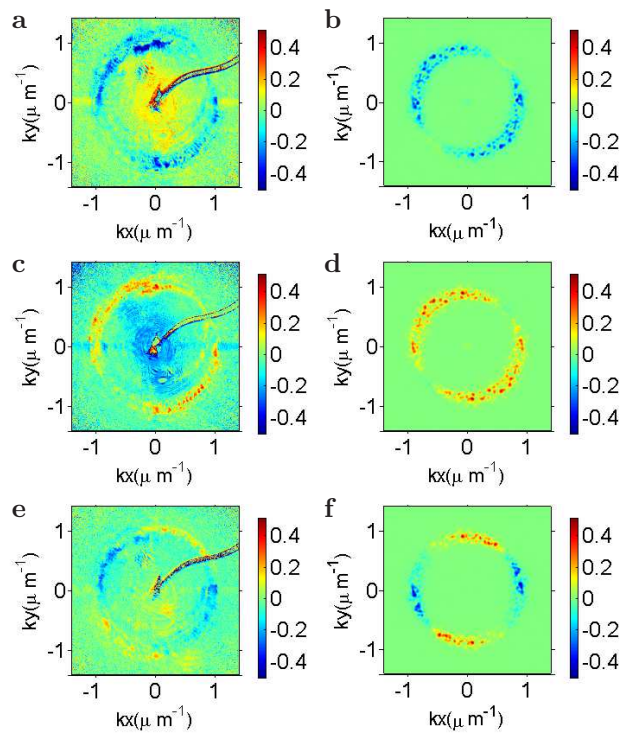}
    \caption{(color online) Far-field circular polarization degree detected
    experimentally (left) compared to theory (right) for the cases
    of horizontal (a, b), vertical (c, d) and diagonal (e, f)
    linearly polarized pump excitation. These three cases correspond
    to Figs.~\ref{fig:Pseudospin}b-d, respectively. Parameters: $l=0.2\mu$m, $W_{rms}=0.05$meV, $E_p=0.63+T(0)$meV,
$\Gamma=0.05$meV, $L=68\mu$m, $\tau_X=100$ps, $\tau_C=2$ps. The polarization splitting arising from anisotropy and TE-TM splitting were each assumed equal to $0.04$meV. The shadow in panels (a), (c) and (e) is a block of the pump field to avoid saturation of the detector.}
    \label{fig:NoDefect}
    \end{figure}

We have considered the anisotropic spin Hall effect
at two different locations on our sample. At the first location (Fig.~\ref{fig:dispsplit}c) the
polariton potential is the typical disordered potential present in
microcavities, evidenced via the meshed pattern in the real space image of the transmission intensity.
In the second location, just $150\mu m$ away from the first one, excitation is performed on top of a photonic defect (Fig.~\ref{fig:dispsplit}d). The presence of the defect results in Airy ring-like patterns which arise from the interference of the $\kk=0$ pumped polaritons and those scattered into the Rayleigh ring. The observed fringe separation is of $\Delta l=5.6\mu m$ which corresponds to the momentum of the scattered polaritons ($k=2\pi/\Delta l = 1.1\mu m^{-1}$).

Figure~\ref{fig:NoDefect} shows, for the first location with shallow disorder (Fig.~\ref{fig:dispsplit}c), the measurement of the circular
polarization degree (z-component of ${\mathbf\rho}$) in momentum space for different pump polarizations. The results are quantitatively reproduced by the numerical solutions of
Eq.~\ref{eq:Shrodinger}.

Under vertical (Fig.~\ref{fig:NoDefect}a-b) and horizontal (c-d) planes of linear polarization of excitation, only two quadrants of circular degree of polarization are visible. This is a direct consequence of the presence of the intrinsic momentum independent effective magnetic field ${\mathbf\Omega_{an}}$, which is 	canceled by the TE-TM field in the positive diagonal direction in the far field (as depicted in Fig.~\ref{fig:Pseudospin}a). Polaritons scattered into this diagonal direction do not feel the action of any field and their linearly polarized pseudospin does not precess, not giving rise to any circularly polarized component. On the other hand, the total field is enhanced in the opposite diagonal direction, 
  giving rise to an increased degree of circular polarization with respect to the case of ${\mathbf\Omega_{an}}=0$.
 
 In the absence of ${\mathbf\Omega_{an}}$, four alternating circular polarization quadrants would be expected as a consequence of the distribution of the TE-TM related magnetic field in the Rayleigh circle, independently of the polarization plane of the injected polaritons~\cite{Leyder2007}, as the characteristic signature of the \emph{standard} OSHE. In the presence of ${\mathbf\Omega_{an}}$ we can recover the four quadrants if we rotate the polarization plane of the excitation with respect to the direction of the intrinsic magnetic field so that they are parallel, as shown in Fig.~\ref{fig:NoDefect}e-f. For such injected pseudospin the composition of ${\mathbf\Omega_{an}}$ and the TE-TM related field are canceled at four points in the Rayleigh circle, which delimit the four quadrants (see Fig.\ref{fig:Pseudospin}d).

    \begin{figure}[t]
        \centering
        \includegraphics[width=8.116cm]{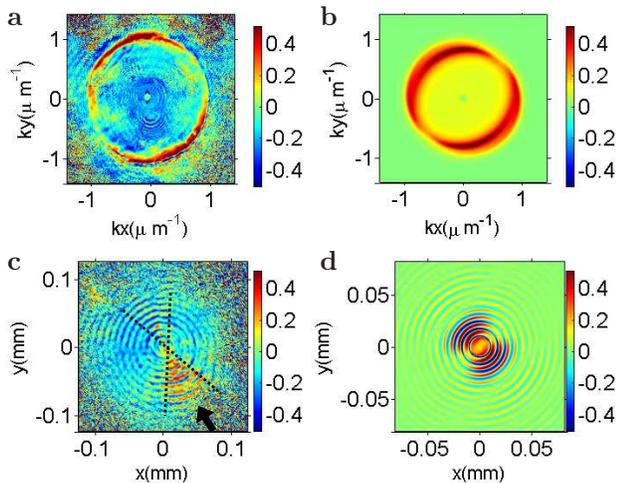}
    \caption{(color online) Far-field [near field] degree of circular polarization detected
    experimentally (a) [c] and compared to theory (b) [d] for the case of
    vertically polarized excitation in the presence of a peak in
    the polariton potential, in the center of panels (c-d).
     The parameters were the same as in Fig.~\ref{fig:NoDefect}, except that the disorder
potential was replaced by a Gaussian shaped defect of height 1meV
and full with at half maximum equal to $3.3\mu$m. Dotted lines in panel (c) show the region where the spin currents flow, in the direction given by the arrow. }
    \label{fig:Defect}
    \end{figure}

We can take advantage of the enhancement in the spin precession related to the existence of the anisotropic field to observe spin currents. They can be directly generated if excitation is performed at the location where the photonic defect is present (Fig.~\ref{fig:dispsplit}d). Here, under vertically polarized excitation, the anisotropic field gives rise a distribution of the circular polarization degree in reciprocal space (Fig.~\ref{fig:Defect}a-b) similar to that of Figs.~\ref{fig:NoDefect}a-b. The presence of the defect results in a strengthening of both the total intensity and the polarization degree (by a factor 2.7) of the Rayleigh signal. The real space images (Fig.~\ref{fig:Defect}c-d) clearly show spin currents of polaritons scattered from the defect in the negative diagonal direction. These spin currents could not be observed in the absence of a defect because in this case scattering takes place homogeneously all over the excitation spot. If we turn by 90$^\circ$ the plane of the polarization of excitation, spin currents follow the same direction with opposite sign (not shown). Note that the concentric ring-like structure of Fig.~\ref{fig:dispsplit}d is still observable in the polarization degree, both in the experimental and simulation images.

\emph{Conclusion.-} The OSHE relies on the combined effects of Rayleigh
scattering of polaritons and pseudospin precession, 
 well described by the introduction of effective magnetic fields physically representing the polarization splitting caused by
TE-TM splitting and/or sample anisotropy. 
 The Rayleigh scattering 
 can
 be engineered to occur at specific points in
the microcavity sample, where directional spin currents are efficiently generated and enhanced due to the presence of the intrinsic anisotropic fields. These features can be exploited for the implementation of integrated light-polarization based logic gates~\cite{Liew2008}.

\emph{Acknowledgements.-} We are very grateful to R. Houdr\'{e} for providing us with the microcavity
sample. This work was supported by the IFRAF. A.A. was funded by the Agence Nationale pour la Recherche, A.B. is a member of the Institut Universitaire de France.


\end{document}